# Extent radiation in different types of radio sources

Alla P. Miroshnichenko[*]

*Institute of Radio Astronomy of NAS of Ukraine*
*4 Chervonopraporna St., Kharkov 61002 Ukraine*

**ABSTRACT**

The contribution of an extent component of source radio emission is estimated for quasars and galaxies. The consideration of source radio structures at kiloparsec scales is used at the decameter and the higher frequency bands. The determination of the contribution of an extent component to source radio emission as well as main physical parameters of sample sources is carried out. We found that especially extent objects, giant radio galaxies, have smaller luminosity of core region, weaker magnetic field and greater characteristic age in comparison with compact radio galaxies and quasars. As it follows from our examination, the extent component contribution to source emission may be the indicator of the radio source age.

## 1. Introduction

It is known that central regions of galaxies and quasars are distinguished by particular intensity at high frequencies. At the same time peripheral extent regions of radio sources are detected best of all at low frequencies. The proportion of radio emission of extent and compact (core) components of sources may characterized by relation of their radio flux densities $S$ at low and high frequencies (for example, at frequencies $\nu$ 25 MHz, 178 MHz, 5 GHz one can use relations $\log(S_{25}/S_{5000})$ or $\log(S_{178}/S_{5000})$). At assumption, that periphery of radio sources is determined by their jets and lobes at kiloparsec scale we can estimate the contribution of the extent component relatively core radio emission for radio galaxies and quasars (Miroshnichenko 2007).

It is of interest to derive the extent component contribution for different types of compact and giant radio sources. For this aim we study radio data for quasars and galaxies at frequencies 25 MHz, 178 MHz, and 5 GHz. In particular, relationships for extent component contribution and redshift $z$, linear size $R$, source age $t_{178}$, magnetic field strength $B$ are considered.

## 2. Extent component contribution and physical parameters of radio sources

To examine the extent radiation of sources we use the VLA observed data for giant radio galaxies (Lara et al. 2001), for radio galaxies of usual linear sizes (Lawrence et al. 1986; Bridle & Perly 1984), and for quasars (Murphy et al. 1993; Reid et al. 1995). The decameter survey of the Northern sky (at the frequency range 10 to 25 MHz) carried out with the UTR-2 radio telescope is the base of our UTR-2 catalogue of extragalactic sources (e.g. Braude et al. 1981, 2003). So, we used the corresponding data at the frequency 25 MHz from the UTR-2 catalogue for the considered radio galaxies and quasars. Besides, we involve necessary data from other catalogues (Kuhr et al. 1981; Veron-Cetti & Veron 1991).

[*]E-mail: mir@ri.kharkov.ua

Thus, we study the sample of giant galaxies, having linear sizes of order of 1 Mpc (79 objects), the sample of galaxies with moderate (standard) linear sizes (76 objects) and the sample of quasars (83 objects). For these sources we calculate the ratio value of corresponding flux densities at low (25 MHz, 178 MHz) and high (5 GHz) frequencies at the logarithmical scale. This value may characterize the extent component contribution relatively core radio emission of galaxies and quasars. Note, that the sample of giant galaxies consists from objects not entered to the zone of the decameter UTR-2 catalogue, and so we can derive only value $\log(S_{178}/S_{5000})$ in this case.

At the decameter band (the frequency 25 MHz) the mean extent component contribution is $<\log(S_{25}/S_{5000})> = 1.72 \pm 0.04$ for galaxies and the mean extent component contribution is $<\log(S_{25}/S_{5000})> = 1.57 \pm 0.08$ for quasars in our samples. These values correspond to the extent component contribution at the frequency 178 MHz (see Table 1) for considered galaxies and quasars.

The analysis of derived extent component contribution at kiloparsec scales and main physical characteristics of given sources is carried out. We calculate the magnetic field strength $B$ and the characteristic age $t_\nu$ of objects with the assumption of synchrotron mechanism of radio emission. So, we assume the equipartition condition for magnetic field energy and energy of relativistic particles (Ginzburg 1979). To estimate the monochromatic luminosity $L_\nu$ and the linear size of objects $R$ we use the flat

Universe model with parameters $q_0 = 0.5$ and $H_0 = 100$ km/s Mpc. Our working formulae are next:

$$B = [48kA(\gamma,\nu)\frac{S\nu}{r\varphi^3}]^{2/7} \; ; \quad (1)$$

$$t_\nu = (\frac{340}{B^3 \nu})^{1/2} \; ; \quad (2)$$

$$L_\nu = S_\nu r^2 (1+z)^{1+\alpha} \; ; \quad (3)$$

$$r = \frac{2c}{H_0}(1 - \frac{1}{\sqrt{1+z}}) \; ; \quad (4)$$

where $k = 100$ (ratio of energies of protons and electrons), $A(\gamma,\nu)$ is the tabulated function (Ginzburg 1979), $\gamma$ is the index of electron energy distribution, $S_\nu$ is the flux density of object at the frequency $\nu$, $r$ is the distance of object, $\varphi$ is the angular size of object; $\alpha$ is the spectral index of radio spectrum, $c$ is the light velocity.

Results of our calculations are presented in Table 1. Namely, we show the mean values of the main physical parameters (the extent radio emission contribution $<\log(S_{178}/S_{5000})>$, the redshift $<z>$, the luminosity of core region $<L_{5000}>$, the magnetic field strength $<B>$, the characteristic age $<t_{178}>$, the linear size $<R>$ of considered objects. Note that in comparison with quasars the extent radio emission contribution has essential value for giant radio galaxies and standard radio galaxies (see Table 1). As it follows from our data, the important physical peculiarity of giant radio galaxies is the low luminosity of their core region. Similar conclusion concerns to value of the magnetic field strength of giant radio galaxies: it is appreciably smaller, than the one for more compact objects - standard galaxies and quasars.

**Table 1.** The extent radio emission contribution and physical parameters of radio sources.

| Objects | $<\log(S_{178}/S_{5000})>$ | $<z>$ | $<L_{5000}>$ (W/Hz) | $<B>$ (Gauss) | $<t_{178}>$ (years) | $<R>$ (cm) |
|---|---|---|---|---|---|---|
| Giant G | $2.60 \pm 0.12$ | $0.158 \pm 0.015$ | $6.45E22 \pm 2.8E22$ | $3.94E-5 \pm 0.97E-6$ | $7.83E6 \pm 6.8E5$ | $2.05E24 \pm 1.54E23$ |
| G | $1.04 \pm 0.04$ | $0.066 \pm 0.012$ | $4.38E24 \pm 2.24E24$ | $3.08E-4 \pm 2.2E-4$ | $7.82E6 \pm 3.27E6$ | $2.54E23 \pm 7.7E22$ |
| Q | $0.73 \pm 0.06$ | $1.13 \pm 0.06$ | $1.31E26 \pm 1.87E25$ | $1.0E-3 \pm 1.0E-4$ | $1.6E5 \pm 2.5E4$ | $3.4E23 \pm 1.8E22$ |

We consider the relationship of the extent radio emission contribution and corresponding redshift, magnetic field strength, linear size, characteristic age for objects in considered samples. Although derived relations display large dispersion, we can note some trends (see Fig.1-6). In particular, for the sample of giant galaxies the extent emission contribution increase when the magnetic field strength decrease (Fig.1). Similar relationship for sample of quasars has greater dispersion (Fig.2). This trend may be due to jet propagation in intergalactic medium, when the essential expansion of source structure drives to weakening of frozen magnetic field.

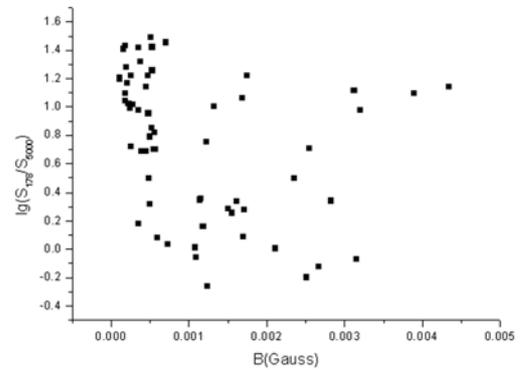

**Figure 2.** Extent component contribution versus magnetic field strength for quasars.

It is of interest, that both sample galaxies and sample quasars display the increase of extent emission contribution at the increase of linear size of objects (Fig.3, 4). Also, extent emission contribution increases when the characteristic age of objects increases (Fig.5, 6).

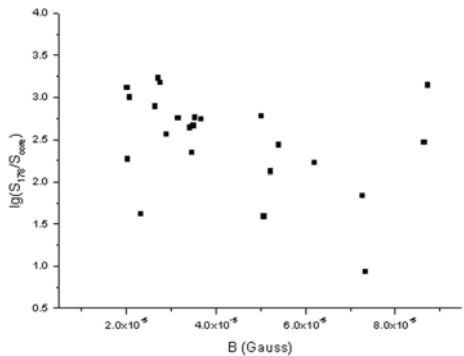

**Figure 1.** Relation of extent component contribution and magnetic field strength for giant galaxies.

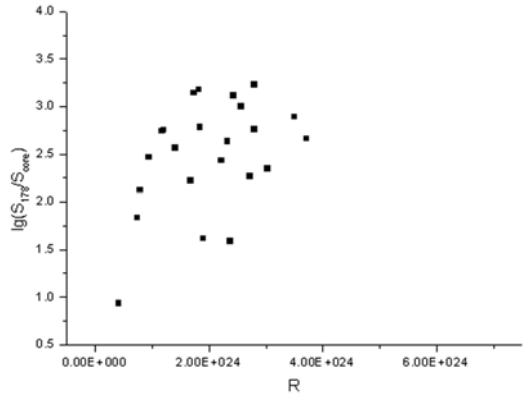

**Figure 3.** Relation of extent component contribution and linear size for giant galaxies.

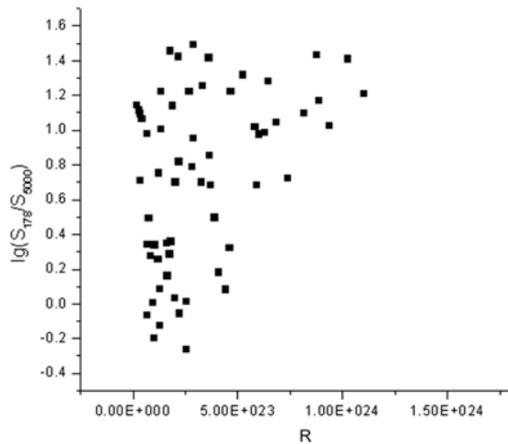

**Figure 4.** Relation of jet share and linear size for quasars.

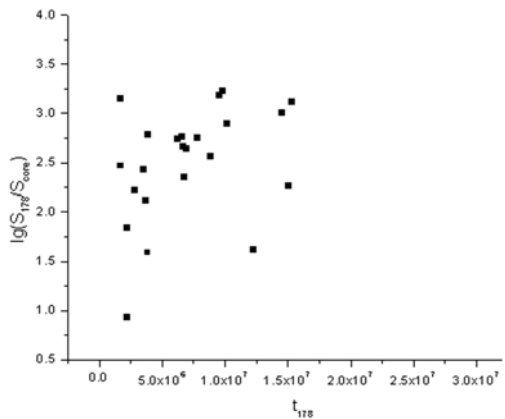

**Figure 5.** Relation of extent component contribution and characteristic age for giant galaxies.

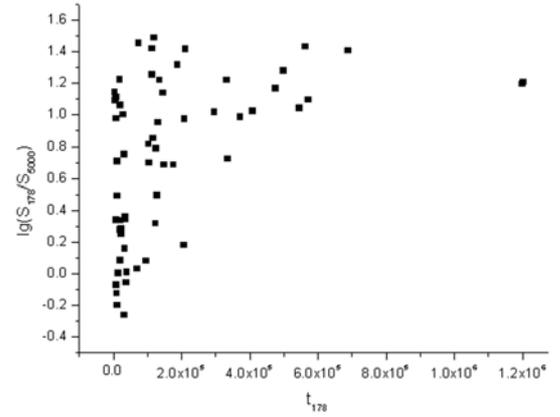

**Figure 6.** Relation of extent component contribution and characteristic age for quasars.

Obviously, the derived relationships are in compliance with the conception of continuous expansion of radio sources at the energy transportation by jet from core region to periphery of radio galaxies and quasars.

### 3. Conclusions

As we derived from the consideration of extent emission in radio sources, the giant galaxies represent the especially great contribution of the extent component to source radio emission. At this, the giant galaxies have smaller luminosity of central region, weaker magnetic field, and greater characteristic age in comparison with compact radio galaxies and quasars.

The linear size of sources is greater at the greater extent emission contribution to source emission.

The magnetic field strength is smaller for radio sources with greater extent emission contribution.

It seems to be important the derived relationship for extent emission contribution and characteristic age of sources. This relationship allows the estimate of the ages of radio galaxies and quasars using the observed radio flux densities at low and high frequencies.